\documentclass{elsart}
\usepackage{amssymb}
\usepackage{graphicx}

\begin{document}

\begin{frontmatter}

\title{On the application of the Critical Minimum Energy Subspace method to
disordered systems}    
\author[CERGY]{Laura Hern\'andez\thanksref{mail1}}
\author[CNEA]{Horacio Ceva\thanksref{mail2}}
\address[CERGY]{Laboratoire de Physique Th\'eorique et Mod\'elisation, UMR CNRS-Universit\'e de Cergy Pontoise,
2 Av Adolphe Chauvin, 95302 Cergy-Pontoise Cedex, France}
\address [CNEA]{Departamento de F{\'{\i}}sica, Comisi{\'o}n Nacional de Energ{\'\i }a At{\'o}mica, Avenida del Libertador 8250, 1429 Buenos Aires, Argentina}   
\thanks[mail1]{Laura.Hernandez@u-cergy.fr}
\thanks[mail2]{ceva@cnea.gov.ar}

\begin{abstract}
We discuss the recent application to strongly disordered systems of the
Critical Minimum Energy Subspace (CrMES) method, used to limit the energy
subspace of the Wang-Landau sampling. We compare with our results on the 3D
Random Field Ising Model obtained by a multi-range Wang-Landau simulation in
the whole energy range. We point out at some problems that may arise when
applying the CrMES scheme to models having a complex free energy landscape.
\end{abstract}

%\pacs{02.70.Tt,02.70.Rr, 64.60.Cn, 75.10.Hk}

\end{frontmatter}

\section{Introduction}

The study of phase transitions via Monte Carlo simulations has recently
regained new interest due to the introduction of the so called "extended
ensemble methods", like multicanonical simulations~\cite{berg}, broad
histogram thecniques~\cite{oliveira}, entropic sampling~\cite{Lee},
Wang-Landau method~\cite{wl1}, among others. These new thechniques increased
significantly the accuracy of the Monte Carlo studies in the cases where
large free energy barriers separate different wells in the free energy
landscape. This step forward in the developement of the Monte Carlo method
is analog to the already largely used clustering methods~\cite{swendsen-wang}%
~\cite{wolff}, which help to overcome the critical slowing down observed at
second order transitions.

Following the arrival of these new methods an avalanche of thechniques were
developped to further improve accuracy and performance of the different
algorithms. Each one of these has been validated through the application to
well known systems, typically those that can be solved exactly like the 2d
Ising model or those whose numerical or aproximated analytical results are
out of discussion like the Potts model for $q\>=5$ (to test the case of
first order transitions).

Recently, Malakis, Peratzakis and Fytas \cite{malakis1} introduced an
interesting approximation method to allow for an easy extension of a
Wang-Landau (WL)~\cite{wl1} study to large systems. It is based on the fact
that in the thermal averages sums at a given temperature T, only some terms
give a relevant contribution: those whose energies correspond to the
interval around the maximum of the probability distribution of the energy at
the corresponding temperature, $P_{T}(E)$. Hence an algorithmic procedure to
determine a restricted interval $\Delta \tilde{E}$ centered in the energy $%
\tilde{E}$ of the maximum of $P_{T}(E)$, called minimum energy subspace
(MES) is proposed. In this way the WL sampling needs to be performed only in
this restricted energy interval, improving the efficiency of the algorithm
and diminishing the errors introduced when joining together the different
parts of the density of states in a multirange simulation~\cite{wl2}.

This algorithm is based on the equivalence of thermodynamical ensembles and
on the central limit theorem: the energy probability distribution at a given
temperature T approaches a Gaussian. Moreover, for a continuous transition,
it is supposed that this remains the case \textit{even at }$\mathit{T_{c}(L)}
$, the critical temperature of a sample whose linear size is $L$. Then one
expects that the width of the critical MES (CrMES) will be of the same order
that the standard deviation of the energy:%
\begin{equation}
\Delta \tilde{E}\propto \sqrt{NT^{2}C}  \label{cmes}
\end{equation}

where volume of the system is given by $N=L^d$, with $d$ the space
dimension, $T$ is the temperature and $C$ the specific heat.

The CrMES is iteratively built, starting from the central value $\tilde{E}$
and extending the interval on both sides of it, until the difference between
the specific heat calculated using the whole energy interval (or the exact
one, if known) and the one calculated using the restricted iterated interval
becomes less than a given error. Assuming that one imposes the same level of
error for all the sizes, a lattice size dependence remains: the center of
the CrMES $\tilde{E}$ and its boundaries are functions of $L$. At the
critical temperature, using the scaling law for the specific heat one gets:

\begin{equation}
\frac{\Delta \tilde{E}}{L^{d/2}} \approx L^{\alpha/2 \nu}  \label{scaling}
\end{equation}

In this way, performing a WL simulation over the whole energy range for
small lattices and determining the CrMES for this small size will be enough
to extrapolate the CrMES where the WL simulation has to be carried out for
larger lattices.

This method has been succesfully applied to the study of the pure 2D and 3D
Ising model.

In this work we are interested in the application of this technique to
highly disorder systems, as has been proposed in~\cite{malakis2} for the 3D
Random Field Ising Model (3D-RFIM) with bimodal distribution of the random
fields. The aim of this work is to show, by the study of the 3D-RFIM, why
the generalization of the CrMES method to highly disordered systems is not
straightforward.

The article is organised as follows in Sec. II we describe the method used
in this work, in section III we present our results and in Sec.IV we discuss
how these results point out to the aspects of the proposed method that have
to be handled with extreme care when applying it to the study of highly
disordered systems.

\section{Description of model and the method}

The 3D-RFIM is one of the simplest disordered systems;\thinspace its
hamiltonian is given by: 
\begin{equation}
\mathcal{H}=-J\sum_{<ij>}s_{i}s_{j}-\sum_{i}h_{i}s_{i}  \label{hamil}
\end{equation}

where $h_{i}$ are the local fields of intensity $h_{0}$ which we assume to
be distributed as follows:

\begin{equation}
p(h_{i})=\frac{1}{2}\left[ \delta \left( h_{i}-h_{0}\right) +\delta \left(
h_{i}+h_{0}\right) \right]  \label{prob}
\end{equation}

The nature of the transition of this model is a subject of controversy since
long ago. In~\cite{nous} we have studied it using the multi-range version of
the WL simulation on the whole relevant energy range. Our results for high
values of the random field intensity show strong first order properties. The
same result had also been found in~\cite{moi} using a completly different
calculation method, namely the canonical Histogram Monte Carlo and similar
features have also been found in the case of a gaussian distribution of
random fields~\cite{machta}

To analyse the method proposed in \cite{malakis2} we start from the density
of states (DOS) calculated using a multi-range WL simulation on the whole
energy space~\cite{nous} and we recalculate thermal averages limiting our
DOS data only to the energy subintervals corresponding to those indicated in~%
\cite{malakis2}.

With this procedure we are able to point out to the different aspects that
should be handled with special care in order to apply this technique to
highly disordered systems. \vspace{0.8cm}

\section{Discussion of results}

The fact that the nature of the transition of the 3D RFIM is still a matter
of controversy calls for a special attention on the validity of the
hypotesis of the methods used to study this model.

Let's discuss first the relationship between the nature of the transition
and the CrMES technique. The method introduced in~\cite{malakis1} is based
on the assumption that one deals with a second order transition. This is
required for Equation~\ref{scaling} and the gaussian-like shape of the $%
P_{T^{\ast }}(E)$ hypothesis to be valid.

Our WL simulation on the whole energy range shows, in agreement with other
works~\cite{malakis2}~\cite{machta}, that for strong disorder, and big
enough lattices, the $C_L(T)$ curves may present a multiplicity of peaks. In
order to estimate the transition temperature $T^*$, we observe that one of
the maxima of $C_L(T)$, located at $T^*$, is associated to a $P_{T^*}(E)$
curve which shows two well separated peaks of equal height. In general, this
corresponds to the highest maximum of $C_L(T)$.

The double peaked $P_{T^{\ast }}(E)$ curves are an indication of the first
order character of the transition, at the transition temperature $T^{\ast }$%
. In very rare cases it has been found that the double peak is a finite size
effect \cite{behringer}. Recently it has been reported that this could be
the case for the 3D-RFIM \cite{malakis3}, though further calculations
involving sizes beyond $\ L=32$ are needed.

In any case the energy probability distribution at the transition
temperature $T^{\ast }$ for a given finite L, cannot be approached by a
single gaussian. In fact its central value $\tilde{E}$ corresponds then to
the \textit{\textbf{minimum}} of $P_{T^{\ast }}(E)$. Moreover, in the case
of first order transitions Eqs.~\ref{cmes} and~\ref{scaling} are not valid,
and the probability distribution may be approached by a double gaussian~\cite%
{binder}.

A modified version of the CrMES method adapted to a system undergoing a
first order transition has been proposed~in \cite{comment}. It is worthwhile
noticing that this work deals with a system presenting geometrically
frustrated interactions but no quenched disorder.

It is clear that when studying systems where the free energy landscape has a
complex structure due to the existence of a quenched disorder, the
possibility of a crossover to a first order transition must be kept in mind.

Moreover finite size effects and sample to sample fluctuations are sensitive
points studying highly disordered systems.

As it is often the case, the signatures of a first order transition are
observed only for large enough sizes. For smaller sizes the transition looks
continuous, as the correlation lenght may easily reach the size of the
system.

\begin{figure}[tbp]
\includegraphics[width=10cm]{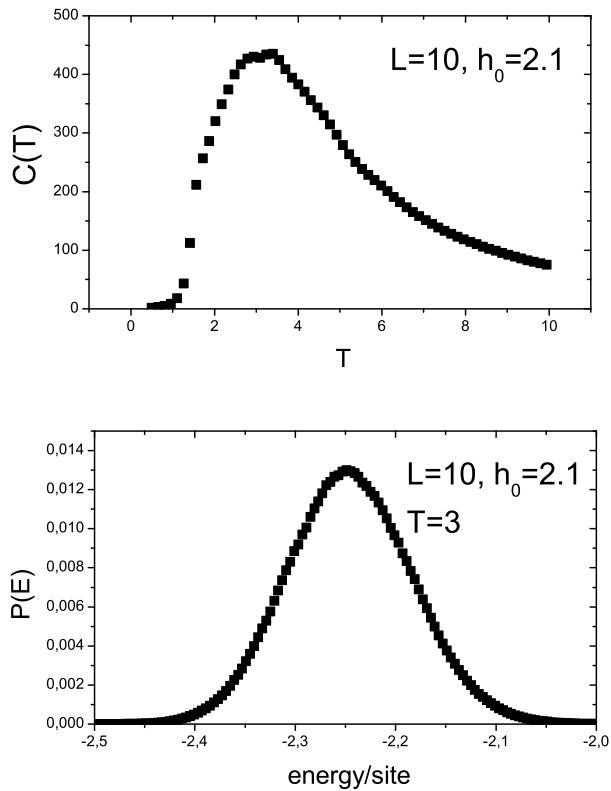}
\caption{$L=10$, $h_{0}=2.1$ (a) Specific heat vs. temperature, (b)
Probability density for T=3, corresponding to the maximum of (a).}
\label{l10}
\end{figure}

Figure~\ref{l10} illustrates this situation for a field $h_{0}=2.1$. In (a)
one can see the specific heat curve for a sample of linear size$\ L=10$ and
in (b) the corresponding $P_{T}(E)$. There is only one peak in $C_{L}(T)$ at 
$T^{\ast }$ and the corresponding probability distribution shows a single
\textquotedblleft gaussian-like\textquotedblright\ peak. The appealing idea
of the CrMES technique should be considered step by step in this case. In ~%
\cite{malakis1} it is proposed to carry WL simulations in the whole energy
range on lattices of (preferably small) size $L$ to determine the CrMES at
this size. Then a new CrMES corresponding to a lattice of size $L^{\prime
}>L $ is calculated via Eq.~\ref{scaling} in order to perform the WL
simulation at this bigger size $L^{\prime }$ only in a restricted energy
interval.

\begin{figure}[tbp]
\includegraphics[width=10cm] {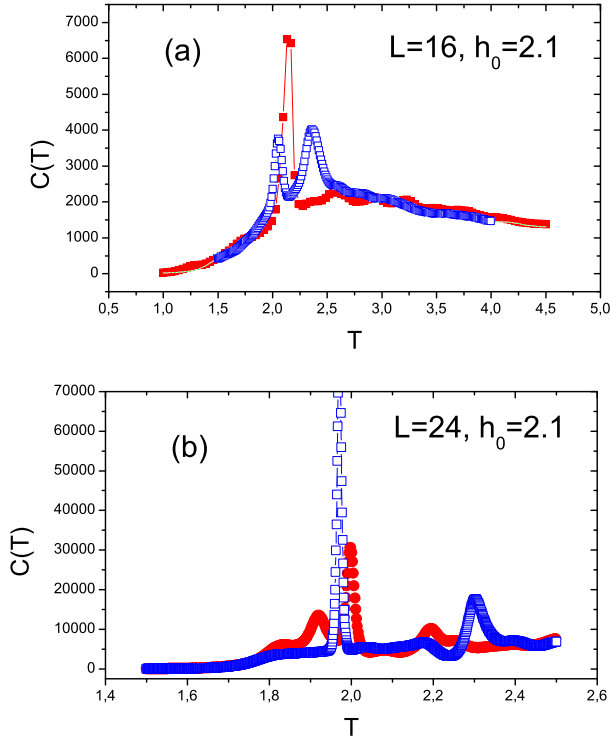}
\caption{$h_0=2.1$ Specific heat vs. temperature, comparison of two
different realizations of quenched disorder for two sizes (a)$L=16$, (b)
L=24 }
\label{shl16l24}
\end{figure}

Figure~\ref{shl16l24} shows that when the size increases new peaks which are
absent for smaller sizes, may appear in the specific heat curve. Moreover
the location of these peaks is strongly sample dependent. In figure~\ref%
{shl16l24} the specific heat curves of two different realisations at the
same field are shown for $L=16$ (a) and $L=24$ (b).

\begin{figure}[tbp]
\includegraphics[width=10cm]{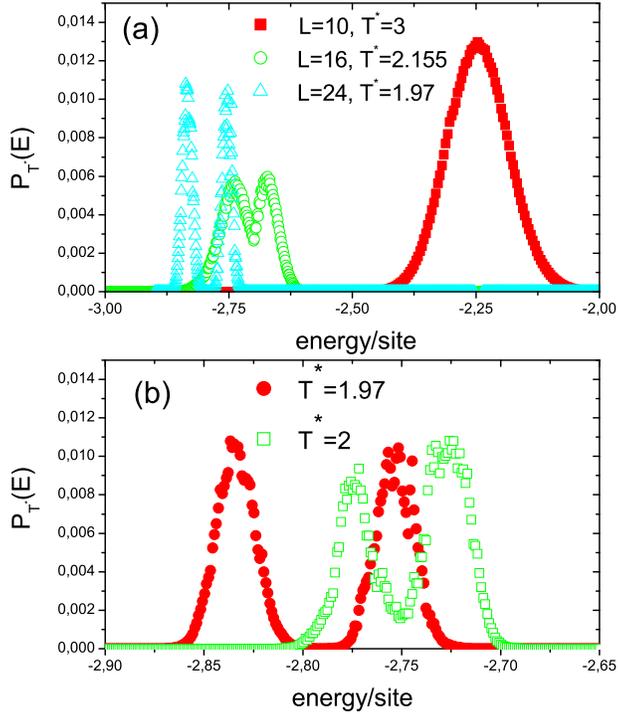}
\caption{(a) Probability distribution of the energy for different lattice
sizes at the corresponding transition temperatures $T^*$. Double peaks
appear in the probability distribution as $L\ge16$ suggesting a first order
transition. Probability distributions of different sizes have very small
overlap. It shows the size of the shift from $\Delta \tilde{E}(L)$ to $%
\Delta \tilde{E}(L^{\prime})$. (b) Probability distribution of the energy
for $L=24$ and two different samples. One can get an estimate of the
extended relevant energy interval for this size.}
\label{probasl}
\end{figure}

Figure~\ref{probasl} shows the energy probability distribution corresponding
to the maxima of $C_{L}(T)$ depicted in Figures~\ref{l10} and ~\ref{shl16l24}%
. It can be seen that for $h_{0}=2.1$, for $L\geq 16$, the probability
distribution of the energy is double peaked. This is a signature of a first
order transition, which is absent for smaller sizes. So the procedure
proposed in ~\cite{malakis1} would be misleading if the starting sizes are
smaller than $L=16$. For the 3D RFIM $L=16$ is big enough so as to render
the use of multirange WL scheme necessary. Very recently a variation on the
WL method has been proposed that migth improve this point \cite{cunha}.

Finite size effects are well known and should be generally taken into
account in order to obtain results for infinite systems. The additional
point when considering finite size effects in disordered systems is that
changing the size implies a change of quenched disorder, which is obviously
not the case in systems without disorder. This becomes particularily
troublesome when large sample to sample fluctuations are observed.

The observed large sample to sample fluctuations are in agreement with other
works on disordered systems~\cite{malakis2}~\cite{machta}~\cite%
{parisi-sourlas}~-~\cite{domany}. As the multiplicity of peaks of the
specific heat and their locations are strongly sample dependent (see Figure~%
\ref{shl16l24}), the location of the maxima of $P_{T^{\ast }}(E)$, and the
resulting CrMES depend also on the considered sample.

To take this aspect into account in~\cite{malakis2} a \textquotedblleft
broadened CrMES\textquotedblright\ for WL simulation is proposed. This
broadened CrMES is determined by the union of M CrMES each of them
calculated for a different quenched random field configuration. Then, using
Equation~\ref{scaling}, the CrMES is estimated for a larger size. It is
assumed that for the new $L^{\prime }>L$, it will be enough to run the WL
algorithm only in this new restricted energy interval.

Letting aside the fact that equations~\ref{cmes} and~\ref{scaling} are not
valid if the transition is first order, and that in general the critical
exponents are not always known in the case of a second order transition, we
will now discuss in detail the idea of extending to a larger size $L^{\prime
}$ the CrMES calculated for $L$~\cite{comment}.

As pointed out above, in the case of disordered systems the sample of size $%
L^{\prime }>L$ is a \textit{\textbf{new, different sample}}, the
corresponding $P_{T^{\ast }}(E)$ may have its maxima located out of the
CrMES that has been extrapolated via Eq.~\ref{scaling}. In Figure~\ref%
{probasl}(a) one can see that there is little overlap between the
probability distributions, even for samples of slightly different sizes. In
addition Figure~\ref{probasl}(b) shows that this is also true for the
probability distributions of two different samples of \textit{the same size}.

We used the DOS obtained in~\cite{nous} for the whole energy range, to
recalculate thermal properties restricting the DOS to the intervals given in
ref~\cite{malakis2}. Let's consider first the case where a broadened
interval is used. For $L=24$ the broadened interval of~\cite{malakis2}
corresponds to the energy interval $\delta E=[-2.91,-1.83]$. We find that,
using this broaden restricted interval, it is possible to reproduce our
results issued from the complete DOS. Unfortunately, though several times
smaller than the whole energy range, this interval is located in the hardest
region from the convergence point of view~\cite{nous}. So the multirange
scheme is unavoidable and several narrow energy intervals are needed to
achieve convergence.

In conclusion, from the efficiency and precision point of view, working with
a very large CrMES guarantees that all the thermodynamical properties will
be correctly reproduced, but the computational effort required is at least
comparable to performing WL in the whole energy range.

The authors of~\cite{malakis2} are aware of this and they have proposed, as
an alternative, to work on the CrMES of \textit{each sample}. To study this
proposal we use our data calculated in the full energy range for each sample 
$i$ and we identify the relevant part of the energy axis for the transition
of the considered sample. This is noted by analogy with \cite{malakis2} $%
\Delta \tilde{E}_{i}$.

Conceptually $\Delta \tilde{E}_i$ corresponds to the region where the energy
probability distribution at the transition temperature is significative for
the sample ${i}$ and, in the case of a double peaked distribution, it
contains the two peaks. Hence, we identify $\Delta \tilde{E}_i$ as the
energy interval where $P_{T^*}(E)$ is different from zero. We then
recalculate the thermal averages using only the DOS restricted to $\Delta 
\tilde{E}_i$. For the sample shown here the $C_L(T)$ curve we obtained using
the whole DOS has two separated peaks. Each of them determines a $T^*$ and
hence a restricted energy interval.

\begin{figure}[tbp]
\includegraphics[width=9cm]{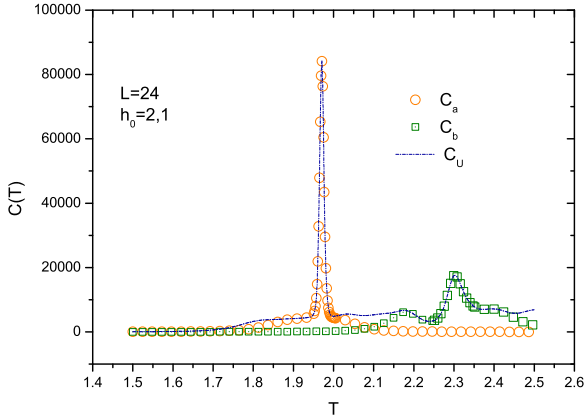}
\caption{(a) Specific heat as a function of temperature $T$, for $L=24$ , $%
h_0=2.1$ obtained from thermal averages using the DOS restricted to three
different energy intervals. Interval $\Delta \tilde{E}_U(L)$ corresponds to
the levels given in~\protect\cite{malakis2} for L=24, this union interval
reproduces a multi-peaked $C_u(T)$ curve coincident to the one obtained with
the whole DOS for the sample $i$. On the other hand the restricted intervals 
$\Delta \tilde{E}_a(L)$ and $\Delta \tilde{E}_b(L)$ give the curves $C_a(T)$
and $C_b(T)$ respectively centered around the temperatures $T_a \approx 1.97$
and $T_b\approx 2.3$. These temperatures correspond to the peaks of the
complete DOS, that gave rise to the restricted intervals, as could be
expected }
\label{int_a_b}
\end{figure}

We show that the whole structure of $C_L(T)$ cannot be reproduced using the
DOS restricted to the interval associated with only one of the peaks. When $%
C_L(T)$ is re-calculated using the DOS restricted to the interval $\Delta 
\tilde{E}_i^a$ ($\Delta \tilde{E}_i^b$) corresponding to the temperature $%
T_a $ ($T_b$) of one of the peaks, the other is not reproduced (see Figure~%
\ref{int_a_b}).

Now let's imagine that we determine the CrMES for a lattice size $L$. 
\textit{As a larger size implies a new different quenched disorder}, it is
possible that the extrapolation to $L^{\prime }>L$ gives a $\Delta \tilde{E}%
^{\prime }$ which doesn't include the highest peak of the new sample $%
i^{\prime }$ of size $L^{\prime }$. In that case the results calculated for $%
L^{\prime }$ will not correspond to the transition region of this new sample.

.

\section{Conclusions}

We show, using the example of the 3D-RFIM, that the CrMES method has to be
applied with extreme care to the study of highly disordered systems, because
the existence of strong quenched disorder may lead to first order transition
properties and strong sample to sample fluctuations.

To begin with, it is worthwhile noticing that when dealing with disordered
systems, increasing $L$ automatically changes the distribution of the
quenched disorder. Hence, if the interval $\Delta \tilde{E}(L)$ calculated
for a sample of size $L$ contains the information to reproduce the highest
peak of $C_{L}(T)$, due to the large sample to sample fluctuations, there is
no warranty that the new, extrapolated interval for $L^{\prime }$ will
reproduce the corresponding peak of $C_{L^{\prime }}$ $(T)$. Then one is
forced to broaden the interval at $L$ to calculate it at $L^{\prime }$. It
should be noticed, however, that there is no rule to control this;
additionally, if the interval is substantially enlarged, the computational
effort is similar to that required when one uses the WL method on the whole
energy range.

Moreover, performing a WL simulation on the whole energy range for small
lattices, in order to determine the nature of the transition before using
the CrMES method should be considered with some care, because it is a
methodology prone to reach incorrect conclusions, as we have shown in Figure~%
\ref{probasl}. As the lattice size increases the two peaks of $P_{T^{\ast
}}(E)$ become well separated, and there is a risk using this method, to
seize only one of them. This migth explain the anomalous behaviour observed
in Fig.5 of~\cite{malakis2} for the largest sizes studied in that work.

The actual nature of the transition of the 3D-RFIM, still under discussion,
does not change the source of the problem. Would the transition be found
second order \ in the thermodynamic limit, the use of \ Eqs. \ref{cmes} and~%
\ref{scaling} would be justified. Nevertheless this would neither change the
problem of sample to sample fluctuations, nor the fact that when working
with finite sizes $P_{T^{\ast }}(E)$ would still be double peaked.

\end{document}